\documentclass{aa}

\usepackage{txfonts}

\newcommand{\teff}{\mbox{T$_{\rm eff}$}}
\newcommand{\logg}{\mbox{$\log g$}}
\newcommand{\monh}{\mbox{\textrm{[m/H]}}}
\newcommand{\vsini}{\mbox{$v\sin i$}}
\newcommand{\loggf}{\mbox{$\log gf$}}
\newcommand{\gamsix}{\mbox{$\gamma_6$}}
\newcommand{\gamfudge}{\mbox{$\Delta\gamma_6$}}

\begin{document}


\title{Spectroscopy Made Easy: Evolution}


\author{N. Piskunov\inst{1} \and Jeff A. Valenti\inst{2}}

\institute{Department of Physics and Astronomy, Uppsala University,
       S-75120 Uppsala, Sweden\\
\email{piskunov@astro.uu.se}
\and
Space Telescope Science Institute, Baltimore, MD 21218\\
\email{valenti@stsci.edu}}

 \date{Received June 15, 2016; accepted August 2, 2016}

\abstract
{The Spectroscopy Made Easy (SME) package has become a popular tool for
analyzing stellar spectra, often in connection with large surveys or exoplanet
research. SME has evolved significantly since it was first described in 1996,
but many of the original caveats and potholes still haunt users. The main
drivers for this paper are complexity of the modeling task, the large user
community, and the massive effort that has gone into SME.}
{We do not intend to give a comprehensive introduction to stellar
atmospheres,
but will describe changes to key components of SME: the equation of state,
opacities, and radiative transfer. We will describe the analysis and
fitting procedure and investigate various error sources that affect inferred
parameters.}
{We review the current status of SME, emphasizing new algorithms and
methods. We describe some best practices for using the package, based on lessons
learned over two decades of SME usage. We present a new way to assess uncertainties
in derived stellar parameters. }
{Improvements made to SME, better line data, and new model atmospheres
yield more realistic stellar spectra, but in many cases systematic errors still
dominate over measurement uncertainty. Future enhancements are outlined.}
{}

\keywords{stars -- spectra -- fundamental parameters -- radiative transfer
               }

\maketitle

%

\section{Introduction}

Strong emphasis on imaging and cosmology over the last 30 years has produced a whole
generation of astronomers with a superficial understanding of stellar spectra.
Recent new directions associated with the discovery of exoplanets, asteroseismology,
and large-scale galactic surveys (e.g., Gaia) has rekindled interest in stars,
with an emphasis on deriving accurate stellar parameters (temperature, mass,
chemical composition, rotation, etc.) in a coherent fashion for a large number of targets.
SME is a \textbf{tool} that we made for exactly this purpose.
We emphasize the word ``tool'' to avoid any misconception: SME helps when used
properly, but as with any tool it can be misused.

SME\footnote{Available at \url{http://www.stsci.edu/~valenti/sme.html}}
has evolved substantially since it was first released nearly 20 years ago.
In the next section we give a short introduction to SME,
followed by a more detailed description of the main components
that have been added or significantly changed since the original version.
These are the the equation-of-state, the continuous and line opacity calculations,
and the radiative transfer solver.
In Section 4 we describe and illustrate with examples
some good practices to use when analyzing stellar spectra with SME.
Section \ref{realistic uncertainties} is dedicated to a methodology for
estimating uncertainties of the resulting stellar parameters.

%
\section{SME under the hood}

SME consists of two components that are loosely connected through
``input'' and ``output'' data structures, that are usually stored in files.
The first component is the graphical user interface (GUI) written entirely
in IDL\footnote{Interactive Data Language, Excelis Visual Information Solutions}.
The GUI helps users create an input structure that specifies the goal of a calculation
and examine the results of a calculation stored in an output structure.
An SME calculation can be a straightforward spectral synthesis
or a more elaborate fit of an observation.
In all cases the user must provide some basic information: global stellar parameters
(effective temperature, surface gravity, metallicity, and radial velocity),
line data, spectral intervals, and line broadening parameters
(macro- and micro-turbulence, $v\sin i$, and instrumental profile).
SME reads line data in the format provided
by VALD\footnote{\url{http://vald.astro.uu.se}} \citep{pis95,kup99}.
Fitting requires additional information:
an observed spectrum and the set of parameters to vary.
The GUI checks the input data for completeness and consistency,
making it a convenient starting point for using SME.

\begin{table*}
\begin{center}
\caption{List of changes to the SME since the original paper \cite{val96},\label{Changes}}
{
\begin{tabular}{| p{3.5cm} | p{5.5cm} | p{8cm} |}
\hline \noalign{\smallskip}
 & ``Old'' SME & Current version \\
\hline \noalign{\smallskip}
Versioning & None & IDL code, library and SME structure \\
\hline \noalign{\smallskip}
Evaluating the number of absorbers & Saha-Boltzmann & Molecular-ionization
                                 equilibrium solver \\
\hline \noalign{\smallskip}
Continuous opacity & Interpolated between the ends of spectral intervals 
                                & Evaluated at the same wavelength points where spectral
                                    synthesis is computed \\
\hline \noalign{\smallskip}
Line opacity & Voigt & Voigt with two-parameter approximation for van der
                                    Waals broadening, new Hydrogen line profile approximation \\
\hline \noalign{\smallskip}
Radiative transfer & Runge-Kutta & Attenuation operator with Bezier splines,
                                spherical radiative transfer \\
\hline \noalign{\smallskip}
Model atmospheres & 2D (\teff - \logg) grid from R.L. Kurucz & 3D (\teff - \logg - \monh) grids from
                                   several sources, new model interpolation techniques \\
\hline \noalign{\smallskip}
LTE & Strict LTE approximation & LTE and NLTE using pre-computed departure coefficients \\
\hline \noalign{\smallskip}
Uncertainties of derived parameters & Using the main diagonal of the approximate covariance matrix &
                              Cumulative distribution estimates in addition to covariance matrix approach \\
\hline
\end{tabular}
}
\end{center}
\end{table*}
Calculations are performed by the second SME component, known as the ``solver''.
IDL code fits observed spectra,
calling a dynamically linked external library to perform the spectral synthesis.
The IDL part reads the input structure and passes all relevant information to the library.
Library functions solve for molecular and ionization equilibrium,
compute continuous and line opacities,
and calculate intensity spectra for specified limb angles.
The IDL code integrates intensities over the stellar disk
and optionally solves for free parameters that yield the best fit.
SME is capable of fitting global stellar parameters
(effective temperature, surface gravity, metallicity, abundances of specified atoms,
$v\sin i$, radial velocity, macro- and micro-turbulence)
and some parameters of spectral lines.
It is important to remember that SME minimizes a weighted $\chi^2$ statistic,
where the usual contribution of each pixel is further weighted
by the observed spectrum.
This formulation gives more weight to points near the continuum
at the expense of points in line cores,
which helps the minimization procedure decouple the influence
of continuum and line parameters.
SME also calculates and returns standard $\chi^2$.

Table~\ref{Changes} lists most of the changes that occurred since the first SME paper.
The following section describe major modifications and enhancements. Since this is more of a ``progress
report'' paper we will not repeat the description of some important SME features, such as the observation
mask, instrumental broadening, $\chi^2$ evaluation, and Marquardt-Levenberg optimization among
others.. More
information can be found in the original paper and in documentation distributed with the package.
We also emphasize SME's compatibility with VALD as a source of atomic and molecular line data.

\subsection{Equation of state} \label{EOS}

Accurate spectral synthesis requires knowledge of the number density
of absorbers and perturbers at every level of the stellar atmosphere. Early versions of SME
(before around 1996) use Saha-Boltzmann equations for that purpose restricting the applications to F and
hotter stars. The latest version relies on a self-consistent equilibrium solver applicable across wide
range of temperature and pressure. The new
equation-of-state (EOS) library functions in SME solve for chemical and ionization equilibrium,
returning number density and partition function for each species that
affects thermodynamic state or contributes to opacities. The output of the EOS is also used
for computing collisional broadening.
The input parameters are temperature, total gas pressure, and atomic abundances.
The EOS provides a consistent foundation for calculations of line and continuum opacities
for cool stars with a variety of species competing for the same atoms.
The equilibrium assumption is appropriate for collision-dominated stellar photospheres.

The EOS solver is based on work by Phil \cite{ben91}.
Let $Z_\mathrm{elem} = n_\mathrm{elem} / n_\mathrm{total}$
be the abundance of a particular chemical element,
where $n_\mathrm{elem}$ is the number density of nuclei of that element
in any form (atomic or molecular, neutral or ion)
and $n_\mathrm{total}$ is the number density of all elements in any form.
Note that the reference is all elements, not just hydrogen.
Let $n_\mathrm{species}$ be the number density
of a particular atomic or molecular species with charge $q_\mathrm{species}$.
Let $X^\mathrm{species}$ be the count of nuclei in a particular species,
that is, 0 for electrons, 1 for atoms or ions, 2 for diatomic molecules, etc.
Let $X^\mathrm{species}_\mathrm{elem}$ be
the count of nuclei of the specified element in a particular species,
e.g.,  $X^\mathrm{H\mbox{$_2$}O}_\mathrm{C}=0$,
$X^\mathrm{H\mbox{$_2$}O}_\mathrm{O}=1$,
and $X^\mathrm{H\mbox{$_2$}O}_\mathrm{H}=2$.
With these definitions, we can write $N$ equations for $N$ elements,
expressing our abundance definitions:
\begin{equation}
Z_\mathrm{elem}=\frac{\sum_\mathrm{species} n_\mathrm{species}X^\mathrm{species}_\mathrm{elem}}
                     {\sum_\mathrm{species} n_\mathrm{species}X^\mathrm{species}}.
\end{equation}
These abundance equations are supplemented
by the particle and charge conservation equations:
\begin{eqnarray}
n_{e^-} + \sum_\mathrm{species} n_\mathrm{species} & = & P/kT\label{Particle conservation} \\
\sum_\mathrm{species} n_\mathrm{species}\cdot q_\mathrm{species} & = & n_{e^-}\label{Charge conservation}
\end{eqnarray}

Eliminating $n_\mathrm{species}$ from Equations (1-3) requires a relationship
between the number density of a species and its constituents.
In ``equilibrium'' the ratio of a reaction product to the reactants depends only on temperature.
This ratio is denoted by $K(T)$ for chemical equilibrium and $I(T)$ for ionization equilibrium:
\begin{eqnarray}
\log n_\mathrm{species} - \sum_\mathrm{elem} X^\mathrm{species}_\mathrm{elem}\cdot\log n_\mathrm{elem}
 & = & \log K(T) \\
\log n_\mathrm{neutral} - \log n_{\mathrm{species}}-q_\mathrm{species}\cdot\log n_{\mathrm{e}^-} & = & \log I(T)
\end{eqnarray}
where $n_\mathrm{neutral}$ is the number density of neutral counterpart of a $q$-times ionized $species$. 
For example, these equations:
\begin{eqnarray*}
  \log n_{\mathrm{H}_2\mathrm{O}} - 2\log n_\mathrm{H} -\log n_\mathrm{O}
  &=& \log K_{\mathrm{H}_2\mathrm{O}} \\
  \log n_\mathrm{TiO}-\log n_{\mathrm{TiO}^+}-\log n_{\mathrm{e}^-}&=&\log I_{\mathrm{TiO}^+}
\end{eqnarray*}
connect the number of water molecules with its constituent atoms and the numbers of neutral and ionized titanium
monoxide molecule with the concentration of free electrons.

Equations (4-5) can be combined to provide explicit expressions for $n_\mathrm{species}$
\begin{eqnarray}
 \log n_\mathrm{species} & = & \log K(T) -\log I(T) + \sum_\mathrm{elem} 
 X^\mathrm{species}_\mathrm{elem}\cdot\log n_\mathrm{elem} \nonumber \\
& & - \log n_{e^-}^{q_\mathrm{species}}
\label{species}
\end{eqnarray}

After eliminating $n_\mathrm{species}$
using Expression (\ref{species}) above that combines chemical and ionization equilibrium
equations (4-5), we are left with only $N+1$ independent variables:
number densities of neutral atoms ($n_\mathrm{atom}$) for $N$ elements
and the number density of free electrons ($n_\mathrm{e}^-$).
This seems to be smaller than the number of equations ($N+2$),
but the sum of all $N$ equations in (1) is unity by definition,
reducing by one the number of constraints restoring the match with the number of unknowns.

The chemical equilibrium constant $K$ can be written as combination of masses, partition
functions, and temperature:
\begin{equation}
K(T)  =  \left(\frac{2\pi kT}{h^2}\right)^{3/2}\cdot 
             \left(\frac{\prod\limits_\mathrm{elem} m^{X^\mathrm{species}_\mathrm{elem}}_\mathrm{elem}}
             {m_\mathrm{species}}\right)^{3/2}
              \cdot\frac{\prod\limits_\mathrm{elem} Q^{X^\mathrm{species}_\mathrm{elem}}_\mathrm{elem}}
             {Q_\mathrm{species}} \cdot e^{-D_\mathrm{species}/kT} \\
\end{equation}
where $k$ is the Boltzmann constant, $m$ is a mass, $Q$ is a partition function,
and $D_\mathrm{species}$ is a dissociation energy.

The ionization equilibrium constant $I$ for two consecutive ionization stages is given by the well-known
Saha equation: 
\begin{equation}
1/I(T)\equiv \frac{n_\mathrm{ion+1}n_{\mathrm{e}^-}}{n_\mathrm{ion}} =
                  \left(\frac{2\pi kTm_{\mathrm{e}^-}}{h^2}\right)^{3/2}
                  \cdot \frac{2 Q_\mathrm{ion+1} }
                  {Q_\mathrm{ion}}
                  \cdot e^{-E_\mathrm{ion}/kT}
\end{equation}
where $E_\mathrm{ion}$ is the ionization energy from ion to ion+1. Combining Saha equations
permits writing the expression of $I$ connecting higher ionization stages with the neutral form.

Currently, the SME library includes partition function for 99 atoms
from Hydrogen to Einsteinium with six or more ionization stages for all abundant elements.
These partition functions were generously provided by R. L. Kurucz
(\url{http://kurucz.harvard.edu/atoms/pf/}).
For molecules we collected partition functions and molecular equilibrium constants
for 197 molecules and six negative ions, mostly from \citet{sau84}.
\citet{bar16} recomputed molecular partition functions for diatomic molecules and a few negative ions,
using updated energy level data from the literature and the NIST database.
Their calculations cover temperatures from 10 K to 10000 K,
suitable even for interstellar medium conditions.
We fit their finely tabulated partition function values with a function
that minimizes the maximum error over the full temperature range.
Table~\ref{List of species} lists the molecular species currently in SME
and the source of each partition function.

\begin{table*}
\begin{center}
\caption{List of molecular species and negative ions included the EOS. The ST and BC in
the Ref. fields indicate the origin of the 
molecular equilibrium constants and partitions functions from Sauval \&\ Tatum or
Barklem \&\ Collet,\label{List of species}}
{
\begin{tabular}{| ll | ll | ll | ll | ll | ll |}
\hline\hline
Species & Ref. & Species & Ref.  & Species & Ref. & Species & Ref. & Species & Ref. &
Species & Ref. \\
\hline
Al$_2$   &ST&Al$_2$O  &ST&Al$_2$O$_2$ &ST&AlBO$_2$ &ST&AlCl  &ST&AlCl$_2$ &ST \\ 
AlClF    &ST&AlF      &BC&AlF$_2$     &ST&AlH   &BC&AlO   &BC&AlO$_2$  &ST \\ 
AlO$_2$H &ST&AlOCl    &ST&AlOF        &ST&AlOF$_2$ &ST&AlOH  &ST&AlS   &BC \\ 
BH$_2$   &ST&BH$_3$      &ST&BO          &ST&BO-   &ST&BO$_2$   &ST&BO$_2$H$_2$ &ST \\ 
BaCl$_2$ &ST&BaClF    &ST&BaF$_2$     &ST&BaO$_2$H$_2$&ST&BaOH  &ST&Be$_2$O  &ST \\ 
Be$_3$O$_3$    &ST&BeBO$_2$ &ST&BeC$_2$     &ST&BeCl$_2$ &ST&BeF$_2$  &ST&BeH$_2$  &ST \\ 
BeH$_2$O$_2$&ST&BeOH  &ST&C$_2$       &BC&C$_2$-   &BC&C$_2$H   &ST&C$_2$H$_2$  &ST \\ 
C$_2$H$_4$  &ST&C$_2$HCl &ST&C$_2$HF     &ST&C$_2$N   &ST&C$_2$N$_2$  &ST&C$_2$O   &ST \\ 
C$_3$       &ST&C$_3$H      &ST&C$_4$          &ST&C$_5$    &ST&CH    &BC&CH-   &ST \\ 
CH$_2$   &ST&CH$_3$      &ST&CH$_3$Cl       &ST&CH$_4$   &ST&CHCl  &ST&CHF   &ST \\ 
CHP      &ST&CN       &ST&CN-         &ST&CO    &BC&CO$_2$   &ST&CS    &ST \\ 
CS-      &ST&CS$_2$   &ST&CaCl        &BC&CaCl$_2$ &ST&CaF   &BC&CaF$_2$  &ST \\ 
CaH      &BC&CaO$_2$H$_2$&ST&CaOH     &ST&ClCN  &ST&CrH   &ST&CrO   &BC \\ 
CrO$_2$  &ST&FeCl$_2$ &ST&FeF$_2$     &ST&FeH   &ST&FeO   &ST&FeO-  &ST \\ 
FeO$_2$H$_2$&ST&H$_2$ &ST&H$_2$+      &BC&H$_2$-   &ST&H$_2$O   &ST&H$_2$S   &ST \\ 
H$_3$+      &ST&H$_3$BO$_3$    &ST&H$_3$O+        &ST&HBO   &ST&HBO$_2$  &ST&HBS   &ST \\ 
HCN      &ST&HCO      &ST&HCl         &BC&HF    &BC&HS    &ST&HS-   &BC \\ 
K$_2$Cl$_2$ &ST&K$_2$O$_2$H$_2$       &ST&KBO$_2$  &ST&KCN   &ST&KCl   &ST&KOH   &ST \\ 
LaO      &BC&LaO$_2$  &ST&Li$_2$O$_2$H&ST&LiBO$_2$ &ST&LiCl  &BC&LiOH  &ST \\ 
MgCl$_2$ &ST&MgClF    &ST&MgF$_2$     &ST&MgH   &ST&MgO   &ST&MgO$_2$H$_2$&ST \\ 
MgOH     &ST&MgS      &BC&N$_2$       &ST&NH    &BC&NH$_2$   &ST&NH$_3$   &ST \\ 
NO       &ST&NO+      &BC&NS          &BC&Na$_2$C$_2$N&ST&Na$_2$Cl$_2$&ST&Na$_2$O$_2$H&ST \\ 
NaBO$_2$ &ST&NaCN     &ST&NaCl        &ST&NaH   &ST&NaOH  &ST&O$_2$    &BC \\ 
OBF      &ST&OCS      &ST&OH          &ST&OH-   &BC&OTiF  &ST&P$_4$    &ST \\ 
PH$_2$   &ST&PH$_3$      &ST&PO$_2$      &ST&S$_2$    &ST&SO    &ST&SO$_2$   &ST \\ 
ScO      &BC&Si$_2$   &ST&Si$_2$C     &ST&Si$_2$N  &ST&Si$_3$   &ST&SiC   &ST \\ 
SiC$_2$  &ST&SiF      &BC&SiH         &BC&SiH-  &ST&SiH$_2$  &ST&SiH$_2$F$_2$&ST \\ 
SiH$_3$Cl   &ST&SiH$_3$F    &ST&SiH$_4$        &ST&SiN   &BC&SiO   &ST&SiO$_2$  &ST \\ 
SiS      &ST&SrCl$_2$ &ST&SrF$_2$     &ST&SrO$_2$H$_2$&ST&SrOH  &ST&TiCl  &ST \\ 
TiCl$_2  $ &ST&TiCl$_3$  &ST&TiF$_2$     &ST&TiF$_3$  &ST&TiH   &ST&TiO   &ST \\ 
TiO+     &ST&TiO$_2$  &ST&TiOCl       &ST&TiOCl$_2$&ST&TiS   &BC&VO    &ST \\ 
VO$_2$   &ST&YO       &ST&YO$_2$      &ST&ZrCl$_2$ &ST&ZrCl$_3$ &ST&ZrCl$_4$ &ST \\ 
ZrF$_2$  &ST&ZrF$_4$     &ST&ZrO         &BC&ZrO+  &ST&ZrO$_2$  &ST & & \\
\hline
\end{tabular}
}
\end{center}
\end{table*}

SME uses a Newton-Raphson approach with Ng acceleration to solve Equations (1-8) iteratively.
Convergence of this method depends critically on having a good initial guess.
SME uses the following elegant and remarkably robust algorithm to obtain an initial guess.
Apportion the input total pressure to atomic species according to their relative abundance.
Use the equilibrium constants to calculate partial pressures  (Eq. \ref{species}) of most abundant species.
Scale the atomic partial pressures by the ratio of the input pressure
divided by the sum of current atomic and molecular partial pressures.
Iterate the cycle above a few times until the scale factor deviates from unity by less than 1\%.
The result is close to the true solution for all physically possible conditions,
even though the procedure ignores coupling between molecular species.
With this initial guess, the Newton-Raphson solver typically achieves a precision
of 10$^{-5}$ in 3-5 iterations for every partial pressure.

\subsection{Line opacities in SME} \label{line opacity}

SME has a new and improved approximation for the atomic hydrogen line opacity (see below),
while Voigt function:
\begin{equation}
H(a, v) = \frac{a}{\pi}\int_{-\infty}^{+\infty}\frac{e^{-x^2}dx}{a^2+(v-x)^2}
\label{Voigt}
\end{equation}
is used for spectral lines of other species. Here $a$ and $v$ here are the Lorentz broadening and detuning
from the central wavelength expressed in units of Doppler width.
Line opacity calculations are the most time-consuming part of spectral synthesis,
so optimization is important. SME computes line opacity in two steps. First, we compute the
central line opacity at each atmospheric layer using the number of absorbers provided by the
EOS, line oscillator strength, and excitation energy of the lower level. At this step we identify the
lines that will have no detectable contribution to the final spectrum and ignore them in subsequent
opacity calculation.
For lines other than hydrogen, we also pre-compute the Voigt function parameter $a$.
In the second step, we multiply the central opacity by the opacity profile
and determine the line contribution interval.
The latter is done by comparing line opacity with continuous opacity
at different wavelength offsets from line center,
We assume a Voigt profile for all non-hydrogen lines.
The Doppler broadening parameter of the Voigt function is a combination of thermal and microturbulent
parts. The microturbulent velocity can be specified explicitly and adjusted by the SME. Three mechanisms
contribute to the Lorentzian broadening of the Voigt profile: natural or radiative damping,
interaction with charged perturbers (quadratic Stark effect), and with neutral perturbers
(van der Waals effect). The latter effect is critical for determining the pressure and thus surface gravity
and typical approximation using power low extrapolation from a broadening value at one temperature is
not good enough for strong transitions with well-developed Lorentzian wings.

\cite{bar00m} computed two-parameter van der Waals approximations for several thousand strong transitions
of elements from Li to Ni, better reproducing line shapes in the low-temperature, high-pressure regime.
Their approach takes advantage of the quantum-mechanical description of collisions with neutral Hydrogen
developed by Anstee, Barklem and O'Mara (ABO, see references in \cite{bar00m}).
The collisional cross-section $\sigma$ can be converted to the broadening of the Voigt profile
for a given collisional velocity $v$ as:
\begin{equation}
\gamma_\mathrm{ABO}=\left(\frac{4}{\pi}\right)^{\alpha/2}\cdot\Gamma\left(\frac{4-\alpha}{2}\right)\cdot
v_\mathrm{ref}\cdot\sigma(v_\mathrm{ref})\cdot\left(\frac{v}{v_\mathrm{ref}}\right)^{1-\alpha}\cdot N_H,
\end{equation}
where $v_\mathrm{ref}$ is the reference velocity value and $N_H$ is number density of neutral Hydrogen.
A power law fit to the dependence on the mean collisional velocity adds the second parameter $\alpha$
to the approximation.
These parameters, computed for many intermediate to strong lines in the solar spectrum,
are available from the VALD database (\url{http://vald.astro.uu.se/~vald/php/vald.php}).
SME readily accepts line data in the VALD format.

Note that the \gamfudge\ fudge factor does not apply to these ``extended'' van der Waals parameters.
You can still use this factor to enhance conventional \gamsix\ values or results of the
Uns\"old approximation when no \gamsix\ data is available.
As theoretical values improve, enhancement becomes less necessary.
 
The new hydrogen line absorption code written by Barklem and Piskunov combines
the self-broadening theory developed by \cite{bar00h},
the unified theory to model the interaction of hydrogen atoms with
charged particles by \cite{vid70} implemented in a code by \cite{kur93},
and the dynamic effects of ions using the model microfield method by \cite{ste94}.

Line opacities are re-evaluated every time the atmospheric structure, abundances or line parameters
are changed although in the last two cases only the opacities for affected transitions are re-computed.  

\subsection{Radiative transfer} \label{radiative transfer}

The radiative transfer (RT) solution is performed using an attenuation operator method with
B\'ezier spline approximation for the source function. This algorithm was selected after many
performance and precision tests for several options including the original SME Runge-Kutta, Feautrier
and Hermitian solvers.
A detailed description of the new algorithm can be found
in \cite{del13}. Here we give a very short summary of the method. In a plane-parallel case the direction
of light propagation is conveniently characterized by $\mu$,
which is the cosine of the angle between the local vertical and the propagation direction.
The specific intensity $I_\lambda$ is then given by the radiative
transfer equation:
\begin{equation}
\mu\frac{dI_\lambda}{d\tau_\lambda}=I_\lambda-S_\lambda \label{RT equation}
\end{equation}
where $\tau_\lambda$ is the monochromatic optical depth measured from the surface into the star and
$S_\lambda$ is the source function. The boundary condition is set deep in the atmosphere where the
intensity can be assumed equal to the local black body radiation. The formal solution of RT Equation
\ref{RT equation} from point $\tau_0$ to $\tau$ is given by:
\begin{equation}
I(\tau) = e^{-\Delta\tau/\mu}I(\tau_0)+\int_{\tau_0}^\tau e^{-(\tau-t)/\mu}S(t)dt \label{RT solution}
\end{equation}
where $\Delta\tau=\tau-\tau_0$. Note, that this equation is invariant with respect to the direction of $\tau$
and can be easily re-written in terms of stellar radius or column mass as independent variables. Assuming
that the source function is changing smoothly along the integration path we can approximate it with an
analytical function and evaluate the integral analytically. The simplest linear approximation requires the
knowledge of the source function at the two ends of the interval. Numerical analysis shows that
higher-order approximations significantly improve precision without refining step size. A quadratic
fit to the source function makes this solver comparable with or superior to the Feautrier method
\citep[see][Fig. 3]{pis02}. A simple parabolic fit will require the value of the source function at three grid
points and it may result in ``overshooting'' - creating unphysical values inside the interval. To avoid this we
used a second order B\'ezier fit to the source function. It achieves high precision on a sparse
grid without over/under-estimating the source function. Algorithmic details can be found in the paper by
 de la Cruz Rodr{\'i}guez \& Piskunov \cite{del13} that also describes the cubic version and even the application to
 the polarized RT.

\begin{figure}
\centering
\includegraphics[scale=0.85]{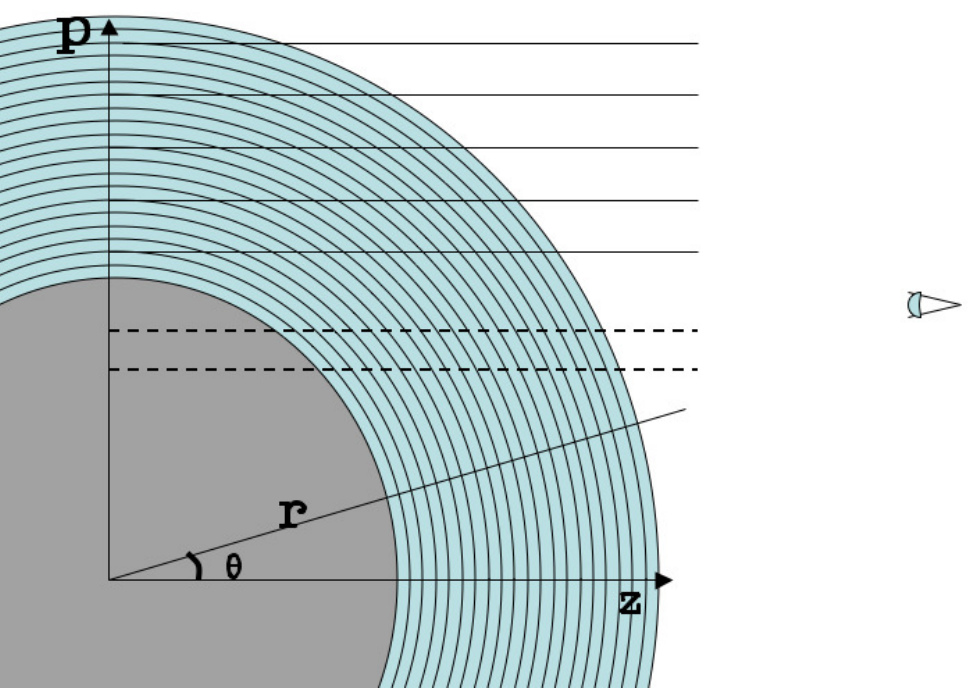}
\caption{
Geometric representation of the spherically-symmetric radiative
transfer problem. The full radius mesh used for representing the variables
as functions of radius is indicated along the radius vector by a set of 
concentric circles. The distance of rays from the stellar center is given
by the impact parameter $p$. The set of rays that do not hit the core are
represented by five parallel rays. The mesh of points used for solving the
equation of radiative transfer are the crossing points between the circles
and the horizontal lines. Distances along the rays are measured by $z = \mu r$
where $\mu = \cos \theta$.\label{Spherical geometry}}
\end{figure}

The plane-parallel case can be generalised to the case of spherical geometry, as presented in
Figure~\ref{Spherical geometry}. In this case, a ray crosses each spherical shell at different $\mu$ angles.
Grazing rays actually cross the atmosphere from surface to surface.
The change in geometry affects the calculation of the optical path from one layer to the next and the solver.
To facilitate solution of RT in a spherical geometry,
SME converts height in the model to impact parameter $p$, relative to the reference radius.
SME now includes and interpolates in a grid of spherical models.  

The attenuation operator algorithm is fast and accurate. Further acceleration comes from an adaptive
wavelength grid described in the original paper. The same algorithm is used to compute the continuum
intensities. SME evaluates continuum at the same wavelength where the line intensity is computed
avoiding continuum interpolation. When constructing the wavelength grid,
SME now takes advantage of the contribution intervals described in Section~\ref{line opacity},
resulting in a substantial performance boost. 

\subsection{Model atmosphere grid interpolation}\label{model interpolation}

SME can use a single atmosphere (e.g., for the Sun), but more commonly SME generates a custom atmosphere
by interpolating in a grid of model atmospheres, based on specified values of \teff, \logg, and \monh.
SME uses the following logic to identify eight ``corner'' models that bracket the desired model.
Find values of \monh\ in the grid that bracket the requested \monh.
Then in this subset of models, find values of \logg\ in the subgrid that bracket the requested \logg.
Finally in this subset of models, find values of \teff\ in the subgrid that bracket the requested \teff.

Next SME pair-wise interpolates (or extrapolates) the eight corner models to produce an output atmosphere.
Pairs of models at each of the four combinations of \logg\ and \teff\ are interpolated to the desired value of \monh.
These four new models are then interpolated to the desired value of \logg,
yielding two models at the requested \monh\ and \logg.
Finally, this pair of models is interpolated to the desired \teff,
producing a single output atmosphere at the specified \teff, \logg, and \monh.
The logic is designed for a regular grid, but handles occasional missing models or changes in grid spacing,
both of which occur in practice.

SME linearly interpolates the logarithm of atmospheric quantities
(temperature, electron number density, atomic number density, mass density)
versus the logarithm of the depth scale.
The depth scale can be mass column or continuum optical depth at a reference wavelength (e.g., 5000 \AA).
By default SME interpolates atmospheres versus continuum optical depth
and solves radiative transfer on a mass column depth scale
because our tests suggest that this often yields better precision.

For each atmospheric quantity, SME fits the first atmosphere in a pair with the second,
allowing a scalar offset in the logarithm of the depth scale and in the logarithm of each atmospheric quantity.
\citet[section 4.1]{val05} illustrate why the depth scale needs to be adjusted before interpolation.
A penalty function is added to the $\chi^2$ goodness-of-fit metric to discourage large offsets in the depth scale.
SME constructs a common output depth scale
by adopting the mean of the offsets determined for each atmospheric quantity.
Both atmospheres are then shifted onto the output depth scale and combined
in proportion to the offset from the grid point to the specified value of \teff, \logg, or \monh.

We evaluated the precision of SME atmosphere interpolation by comparing each atmosphere in the grid
with the atmosphere obtained by interpolating the two adjacent atmospheres along one parameter axis.
In these tests, the grid spacing is effectively doubled along one axis, yielding larger errors than during normal operation.
We calculated the maximum fractional error (``maximum error'') in temperature at any depth in the atmosphere
and the mean absolute value of the fractional error (``mean error'') in temperature at all depths.
For solar-type atmospheres in an ATLAS9 grid, the maximum errors are
1.4\% for interpolation along the \teff\ axis with a grid spacing of 250 K,
0.1\% along the \logg\ axis for 0.5 dex spacing, and 0.2\% along the \monh\ axis with 0.1 dex spacing.

The SME distribution includes atmosphere grids generated with MARCS \citep{gus08}, ATLAS9 \citep{cas03},
and a line-by-line code derived from ATLAS9 \citep{shu04}.
Some of the grids are segregated into separate files based on microturbulence, grid spacing,
or atmosphere geometry (plane parallel vs. spherical).
The suite of available atmosphere grids is described in the SME Manual
and associated documents distributed with the software.

\begin{figure*}
\includegraphics[scale=0.8]{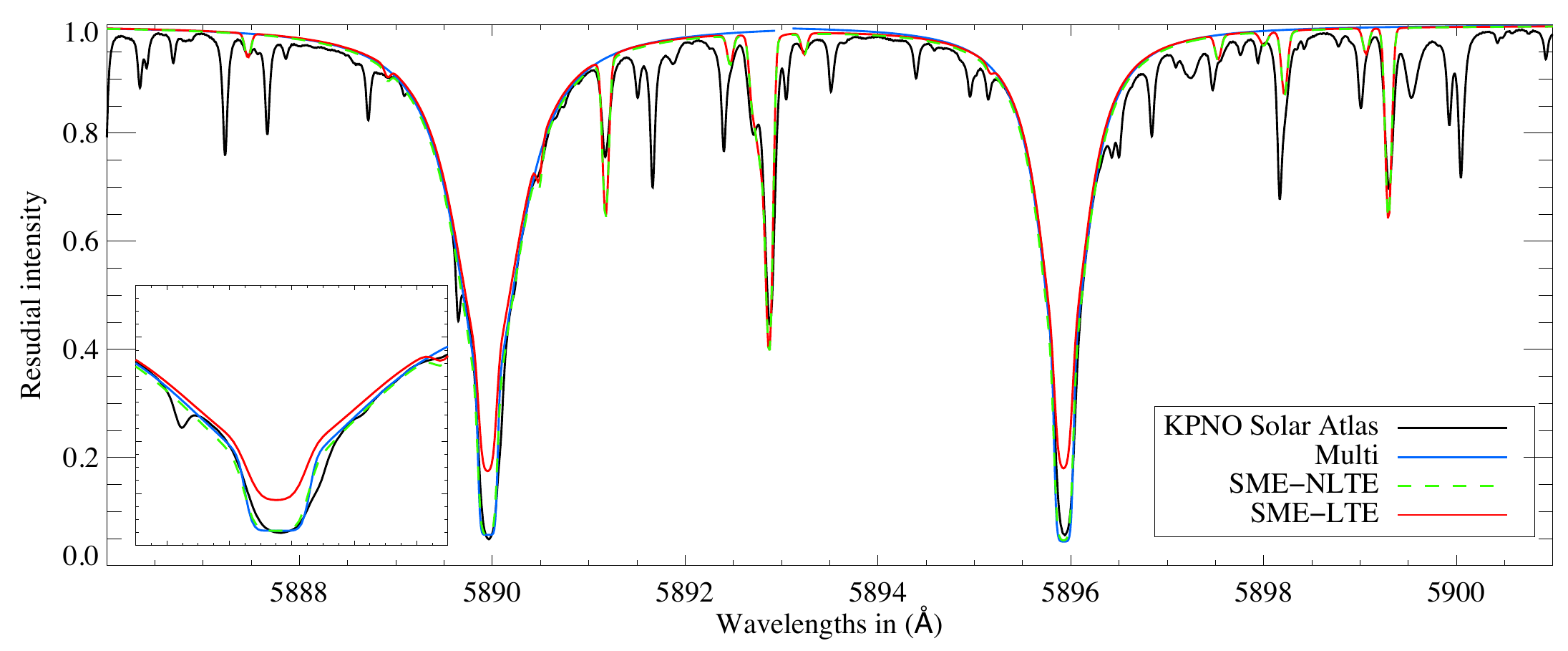}
\caption{Example of NLTE spectrum synthesis for the Sun. The SME calculations used
departure coefficients for sodium precomputed for the MARCS atmosphere
grid and interpolated to the solar \teff, \logg, and \monh.
The MULTI calculation was done separately for each sodium line using a MARCS model
for the Sun. For illustration we also include the observed solar flux spectrum and the SME
LTE synthesis. The insert shows a magnified view of the core of the 5890~\AA\ line.
\label{NLTE synthesis}}
\end{figure*}

\section{NLTE calculations with SME}

SME can account for non-local thermodynamic equilibrium (NLTE) effects on stellar parameters
and individual abundances by applying departure coefficients obtained
by interpolating in externally generated grids of departure coefficients.
Departure coefficients are pre-calculated for all relevant energy levels of the specified element,
for every layer of every atmosphere in the specified atmosphere grid,
and for a range of abundances for the species in question.
SME needs departure coefficients for a range of abundances
because elemental abundances may deviate from the solar pattern.

SME has been used with departure coefficients for Li, C, O, Na, Mg, Al, Si,
Ca, Fe, Ba, and Eu  \citep[e.g.,][]{ber14,oso15,gru16,nord16}.
The current distribution includes grids for Li \citep{lind09},
O \citep{sit13}, Na \citep{lind11,mash08}, Ca \citep{mash08},
Fe \citep{lind12}, and Ba \citep{mash08}.
SME documentation describes the format of the grids and the distribution includes a tool for
generating grids from user departure coefficient data.

The use of pre-computed departure coefficients in SME requires
associations between each transition and specific energy levels.
To make these associations, we use species names,
total angular momentum quantum number J, and term designation.
These data are all available from VALD using long format extraction.
SME still supports the VALD short format for LTE calculations.
SME currently uses a file naming convention to associate
each departure coefficient grid with the corresponding model atmosphere grid.
Once the user selects a model atmosphere grid,
the GUI identifies which transitions can be treated in NLTE
and allows the user to activate NLTE treatment for the corresponding species.

For completeness, we describe here how the radiative transfer solver incorporates
departure coefficients into line opacities and the source function.
SME computes the continuum in LTE, which seems to be a good approximation for cool stars,
but may be a limiting factor for hot stars.
Departure coefficients $b_i$ are defined as the ratio of NLTE level populations
to LTE level populations, $b_i(\tau) = {n_i^\mathrm{NLTE}(\tau)}/{n_i^\mathrm{LTE}(\tau)}$.
The line opacity is then computed as:
\begin{equation}
\kappa_\mathrm{line}^\mathrm{NLTE}=\kappa_\mathrm{line}^\mathrm{LTE}\frac{b_l e^{h\nu/kT} - b_u}
  {e^{h\nu/kT}-1}
\end{equation}
where $\kappa_\mathrm{line}^\mathrm{LTE}$ is the LTE line opacity and $b_l$ and $b_u$ are the
departure coefficients of the lower and the upper energy levels. The corresponding change to the
line source function is:
\begin{equation}
S_\mathrm{line}=\frac{2h\nu^3}{c^2}\frac{b_u}{b_l e^{h\nu/kT}-b_u}
\end{equation}
and the total source function (ignoring scattering) is:
\begin{equation}
S= \frac{S_\mathrm{cont}\kappa_\mathrm{cont}+\sum S_\mathrm{line} \kappa_\mathrm{line}}
{\kappa_\mathrm{cont}+\sum\kappa_\mathrm{line}}
\end{equation}
Figure~\ref{NLTE synthesis} shows the comparison between NLTE SME spectral synthesis based
on interpolation of departure coefficients and the synthetic spectra computed with the MULTI code
\citep{carl86}.

\section{Using SME to derive stellar parameters}

The use of SME to derive stellar parameters involves several steps.
\begin{enumerate}
\item Define spectral intervals that can adequately constrain the models.
\item For each spectral interval, obtain atomic and molecular line data from, e.g., VALD.
If the stellar sample spans a wide range of temperatures, use VALD \textit{extract stellar} mode
to obtain line data at multiple characteristic temperatures.
\item Verify atomic or molecular line  parameters, using the spectrum of a well-characterized
star (e.g., the Sun) as a constraint.
\item Compute an initial synthesis, solving for radial velocity and, perhaps, continuum normalization.
The current version of SME applies continuum adjustments to the observation, rather than the model.
\item Use the GUI to adjust the mask, marking continuum regions that will be used to renormalize
observations, line regions, which will be used to constrain model parameters, and bad regions, which
will be ignored in the fit.
\item Solve for global parameters (e.g., \teff, \logg, \monh), keeping individual abundances fixed.
\item Solve for individual abundances, keeping global parameters fixed, unless certain that
the spectral intervals can adequately constrain global parameters and individual abundances
simultaneously.
\end{enumerate}
\noindent
The details of how to perform these steps are described in the SME documentation.

SME can solve for various model parameters by fitting an observed spectrum.
The stellar parameters are effective temperature (\teff), surface gravity (\logg),
metallicity (\monh), individual elemental abundances, microturbulence,
macroturbulence, projected equatorial rotation velocity (\vsini), and radial velocity.
The atomic and molecular parameters are individual oscillator strengths (\loggf),
individual van der Waals damping parameters (\gamsix),
and a global scale factor for van der Waals damping (\gamfudge).

Finally, SME can fit some properties of the observations, such as continuum level
and spectral resolution.

Individual model parameters can be well or poorly constrained, depending on the
information content in the observed spectrum and physical properties of the star.
Subsets of model parameters may be partially or completely degenerate with each other.
Spectral line diversity is a key factor, but spectral resolution and signal-to-noise ratio are
also relevant. Effective use of SME requires careful assessment of constraints in the
observed spectrum, when deciding which SME parameters to determine. The sensitivity to
specific free parameters (or the lack of) can be verified by trying different initial guesses,
preferably bracketing the final solution.

\begin{figure*}
\centering
\includegraphics[scale=0.081,angle=0]{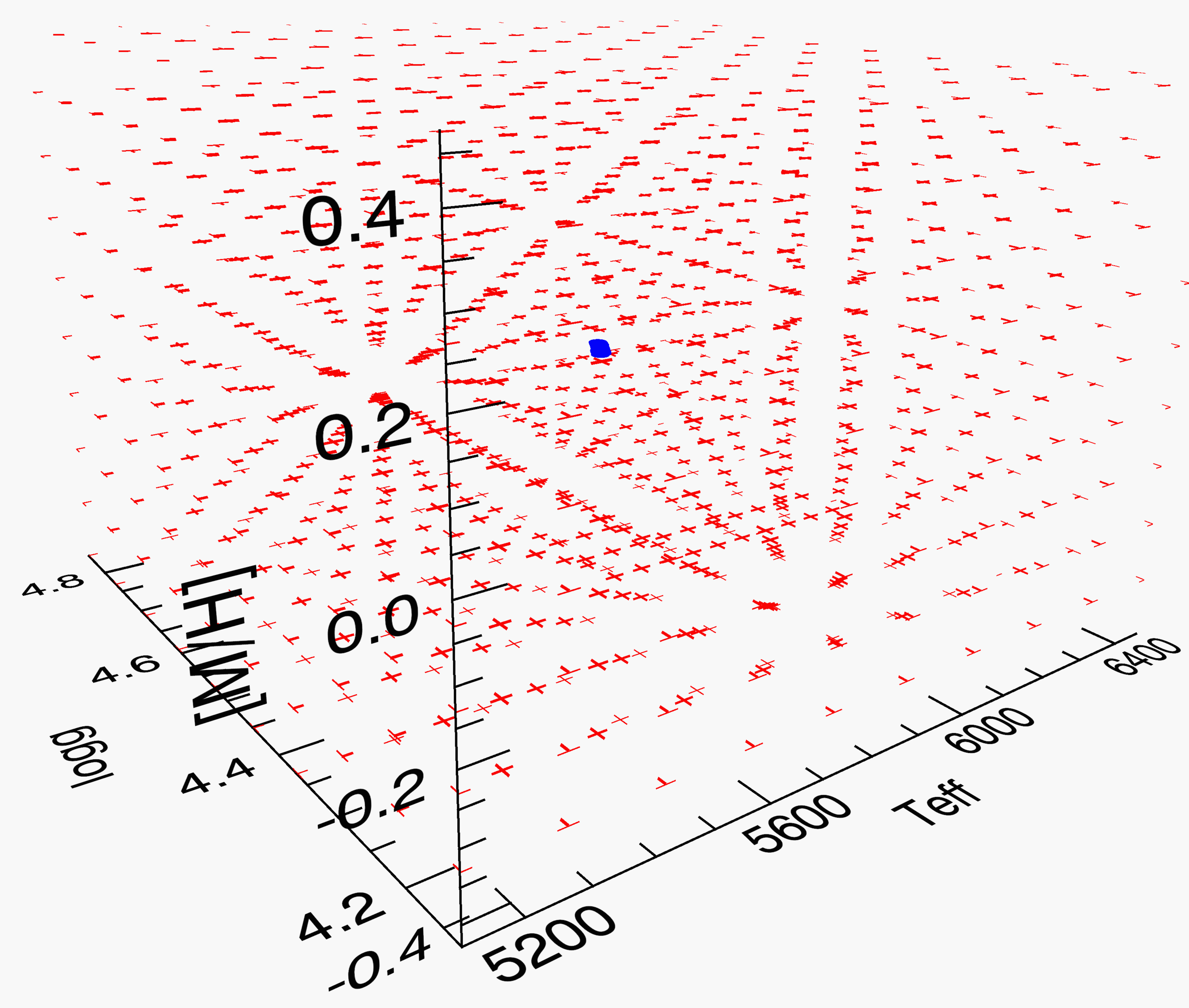}
\includegraphics[scale=0.081,angle=0]{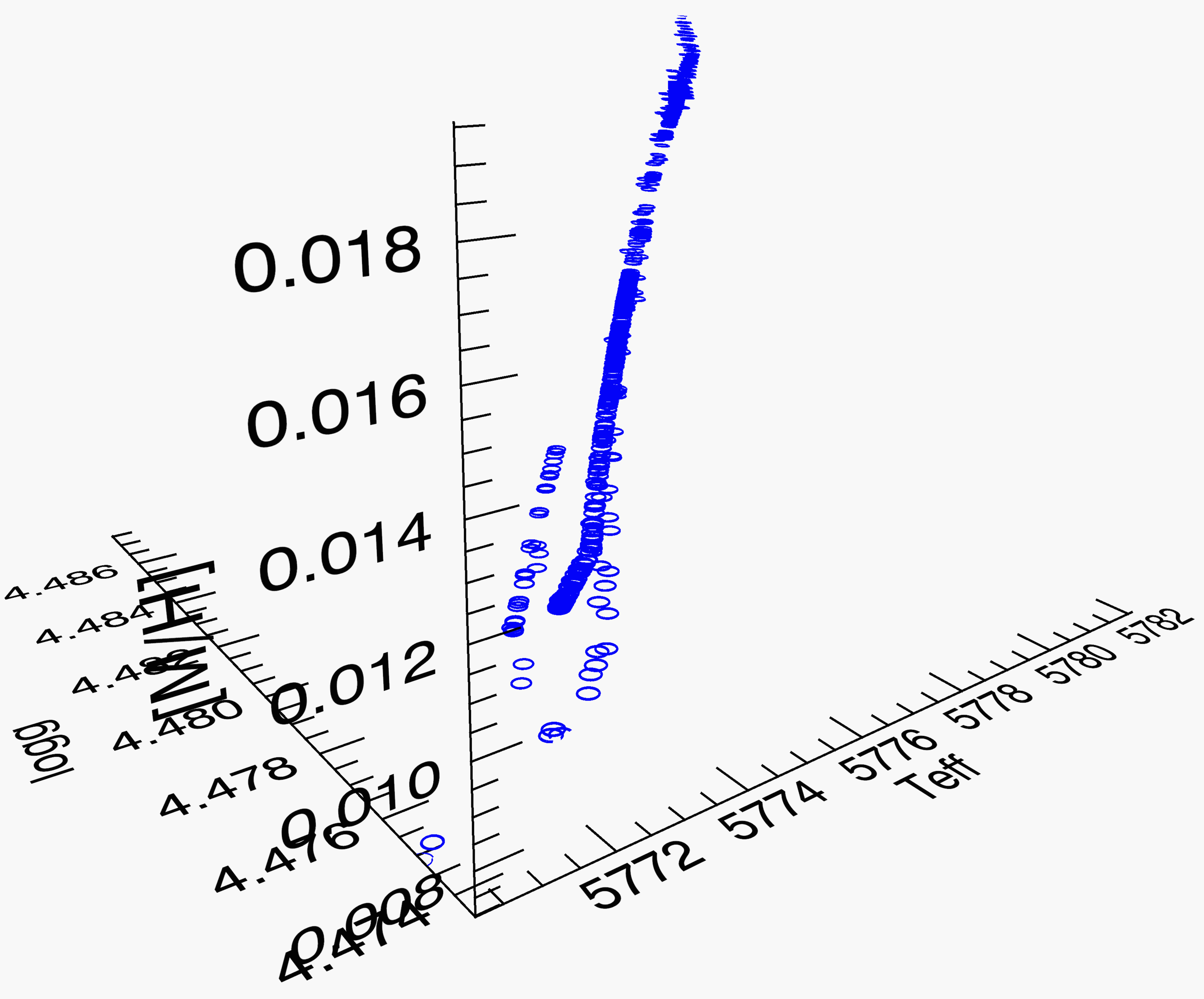}
\caption{
Thousand guesses test, using \object{Vesta} observations with the Keck HIRES
spectrometer. The axes are effective temperature, surface gravity and
metallicity. On the left panel crosses show the initial guesses covering 1200 K
in \teff, 0.5 in \logg\ and 1 dex in metallicity. A tight clump of circles
slightly above the panel center shows the results illustrating robustness of SME
optimization. The right panel shows a close up of 
the results. The spread is very small but has a structure reflecting
cross-dependence of parameters and deficiency of numerical derivatives.
\label{thousand guesses test}}
\end{figure*}

Stellar parameters can be grouped into a subset that mainly affects spectral line strength
(\teff, \logg, \monh, abundances, microturbulence) and a subset that mainly affects line shape
(macroturbulence, \vsini). Instrumental resolution also affects line shape.
Degeneracies tend to be stronger within these subsets and weaker across the two subsets.

Standard spectroscopy texts \citep[e.g.,][]{gra08} discuss how different types of spectral lines
respond to changes in stellar \teff, \logg, and abundance. SME can only disentangle these physical
parameters if the observed spectrum includes adequate constraints. For example, \citet{val05}
use the damping wings of the \ion{Mg}{1} b triplet to constrain \logg\ in main-sequence stars
cooler than $\teff=6200$ K. The damping wings disappear at higher temperature (due to
Mg ionization) or lower gravity (due to lower photospheric density), limiting the effectiveness
of this gravity diagnostic outside the domain where it was originally used.
Ionization balance provides another constraint on photospheric density and hence gravity.
Including spectral lines from neutral and ionized species helps SME constrain gravity
in warmer or more evolved stars \citep[e.g.,][]{bre15}.

SME uses \monh\ to interpolate in an atmosphere grid and also to scale all elemental abundances.
When SME determines \monh\ by fitting a spectrum, the result is a compromise between
the value that yields the best atmosphere and the value that yields the best scaled abundance
pattern. In principle there should be no difference, but in practice SME and the atmosphere grids
likely have slightly different assumptions and errors.

Care is required when solving for \monh\ and individual abundances of selected elements.
Effectively, \monh\ becomes the abundance scale factor for any remaining elements constrained
(perhaps poorly) by the observed spectrum. For example, solving for individual abundances
of some iron-peak elements (V, Cr, Mn, Fe, Co, Ni) while using \monh\ to determine scaled
solar abundances for some $\alpha$ elements (O, Ne, Mg, Si, S, Ar, Ca, Ti) would yield
an \monh\ that
is formally inconsistent with metallicity as defined in $\alpha$-enriched atmosphere models.
A more consistent approach would be to solve for individual abundances of $\alpha$ elements,
while using \monh\ to determine scaled solar abundances for iron-peak elements.

For slowly rotating stars, distinguishing macroturbulence from \vsini\ is difficult,
especially when observed spectral lines are asymmetric due to surface granulation.
In this case, it may be necessary to assume a value for one of these parameters.
For example, \citet{val05} set $\vsini=0$ and determined ``macroturbulence'' for many stars.
Most stars had some contribution from rotation, but a few had negligible rotation.
This lower envelope defined a relationship between macroturbulence and \teff.
For each star in the sample, they used this relationship to tie macroturbulence to \teff\
and then solve only for \vsini.

\begin{figure*}
\centering
\includegraphics[scale=.75,angle=90]{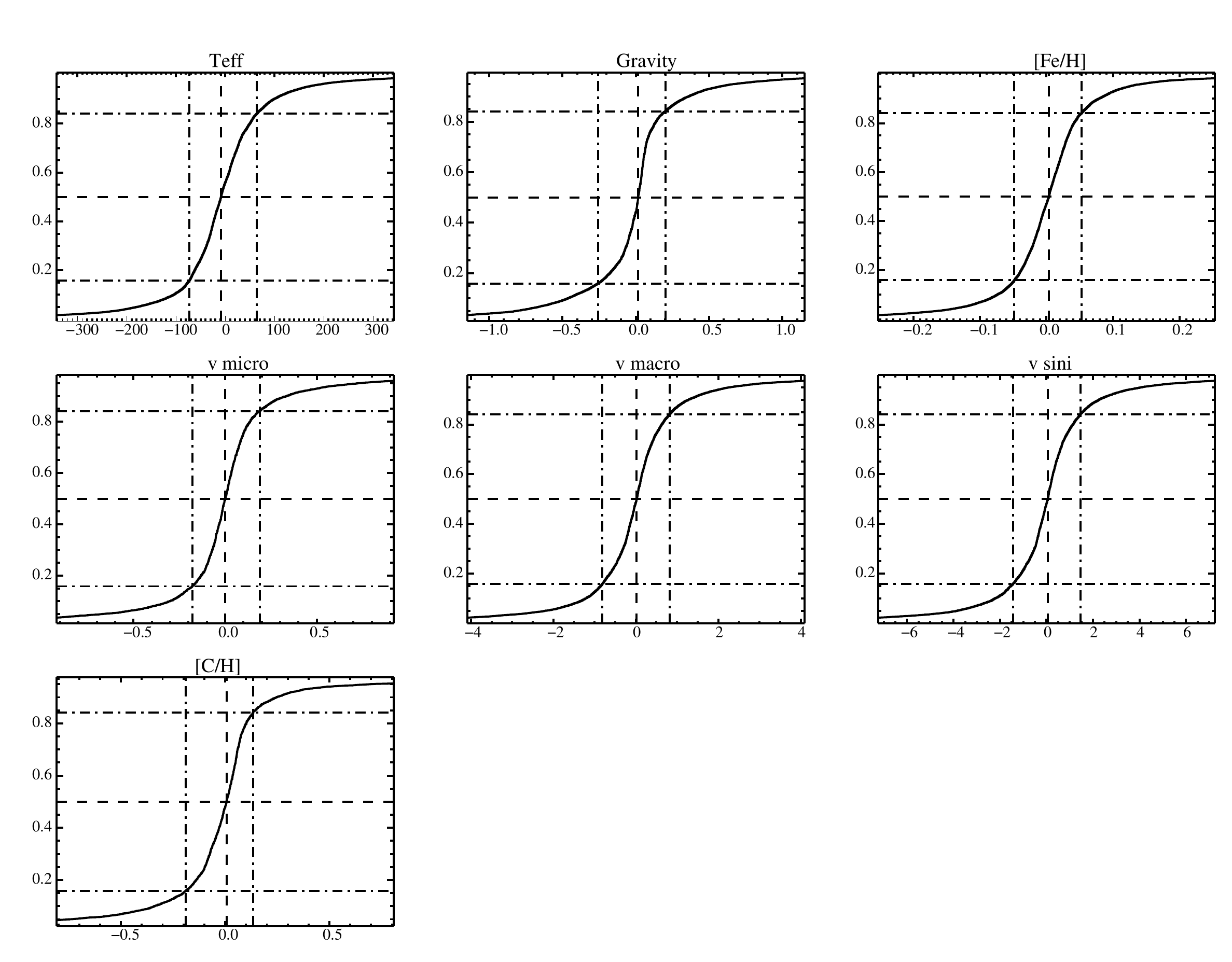}
\caption{
Examples of normalized cumulative distribution functions of parameter offsets
$\Delta p_i$ for an SME fit. The x--axis is parameter offset $\Delta p$ relative
to the value returned by SME. The horizontal dash and dash-dotted lines show
the median and the $\pm 1 \sigma$ levels.\label{parameter uncertainties}}
\end{figure*}

External constraints can help breaking spectroscopic degeneracies. For example, \citet{soz07}
forced \logg\ to be consistent with stellar structure models, constrained by the ratio of
semi-major axis to stellar radius inferred from the light curve of a transiting planet. 
\citet{val09} forced \logg\ to be consistent with stellar models, constrained by spectroscopic
parameters from SME, apparent visual magnitude, and a precise parallax. External
constraints on \logg\ are particularly useful because gravity has a subtle effect on spectra
(see, for example, the comparison of various methods by \cite{bru12}).

Several applications of SME focused on strictly differential analysis of similar stars.
In the case of solar twins, such a comparison can be simplified by selecting the best-fitted
spectral regions and tuning the corresponding line data (oscillator strength and van
der Waals broadening) using solar spectra. This hides inconsistencies in the data and
the model shifting the derived parameters into the reference frame of the Sun. Of course,
the downside of such approach is that accuracy of the results will progressively decrease
for objects less similar to the Sun.

Since SME fits the synthesis to the observations in individual pixels there is no obvious
way to derive microturbulent velocity as in the curve-of-growth analysis: by forcing the
abundance to be independent of of reduced equivalent width. This condition can be checked
after the SME fit, though. Using a number of lines with good parameters covering
a significant range in equivalent width in SME fit will usually come close to this condition.

\section{Deriving realistic uncertainties}\label{realistic uncertainties}
When assessing uncertainties in derived parameters, one must distinguish between numerical
errors in the solver, measurement errors in the data, and physical errors in the model.
For observed spectra of good quality, physical errors in the model typically dominate
other sources of error, yet formal uncertainties returned by the Levenberg-Marquardt
assume measurement errors in the data dominate.
In particular, our formal uncertainties are the square root of elements on the main diagonal
of the covariance matrix, which is the inverse of the curvature matrix \citep{pre02}.
These formal uncertainties typically underestimate the true error ($\chi^2\gg1$),
so SME also returns a heuristic uncertainty estimate that considers model errors.

\begin{figure}
\centering
\includegraphics[scale=0.55]{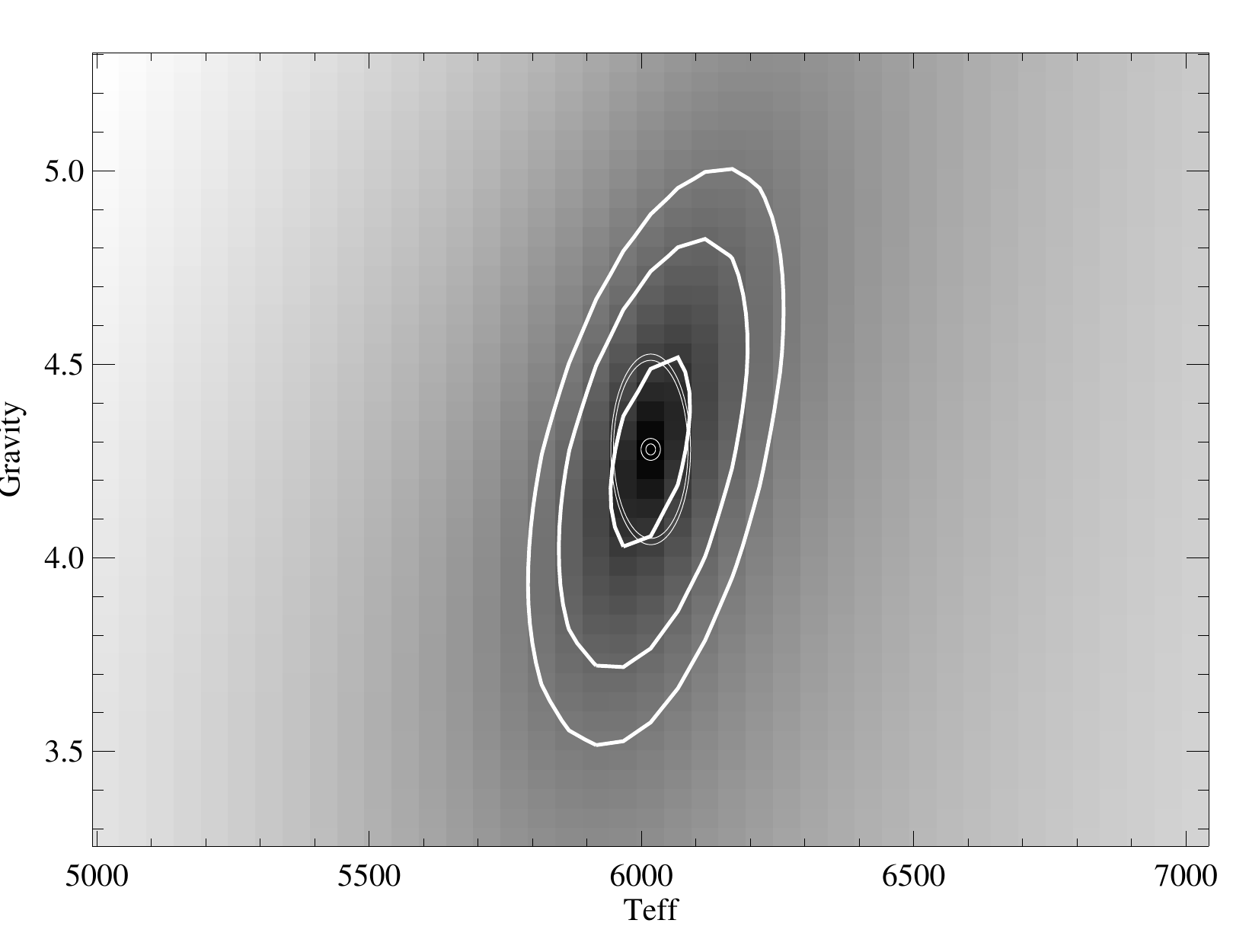}
\caption{
Slice of the $\chi^2$ space along the $T_\mathrm{eff}$-$\log g$ plane for the 
example presented in Figure~\ref{parameter uncertainties}. Contours show 1, 2 and
3 $\sigma$ levels above the minimum determined with procedure described in 
Section~\ref{realistic uncertainties}. The double circle in the center shows the
uncertainties given by the main diagonal elements of the covariance matrix
constructed by SME at the end of iterations. A larger double ellipse comes from
the new uncertainty estimate. \label{chisq surface}}
\end{figure}

Before describing our algorithm for estimating model errors,
we first demonstrated that numerical errors due to the SME solver are negligible.
Components of the solver that can produce numerical errors include
the minimization algorithm, calculation of numerical derivatives,
interpolation algorithms, and ultimately numerical precision.
As a straightforward test,
we used SME to recover stellar parameters for 1000 simulated solar observations,
all based on a single spectrum generated by SME.
We set instrumental resolution to 100,000
and added 0.5\% noise to each simulated observation.
In this case, the SME model is perfect by construction.
For each realization we adopted a random (uniformly distributed) initial
guess for the free parameters.
The initial guesses spanned 1200~K in \teff, 0.5 dex in \logg, and 1 dex in \monh.
The distribution of final values had a standard deviation of 1~K for \teff\
and 0.001 dex for the other two free parameters.
This precise convergence from a wide range of initial values
demonstrates that the SME solver is robust.
However, the final parameter errors are much smaller than errors
encountered when analyzing actual spectra.

Next we used SME to fit an actual solar spectrum
obtained by observing the asteroid Vesta with the Keck HIRES spectrometer.
We used an FTS solar atlas  \citep{kur84} to normalize the stellar continuum.
Again we used SME to recover stellar parameters,
this time starting from a uniform grid of initial guesses for the free parameters.
The distribution of final values had a standard deviation
of 12~K for \teff\ and 0.01 dex or less for \logg\ and \monh.
This is an order of magnitude larger than the previous test.
If the spectral intervals poorly constrained the free parameters,
then degeneracies and parameter errors would be even larger.
Figure ~\ref{thousand guesses test} illustrates the initial and final values
of the free parameters.
The final values are mostly coalesced into a primary sequence,
but two alternate sequences indicate adjacent local minima.
The extent of the parameter sequences indicate partial degeneracy
between the free parameters, which is apparently exacerbated by model errors.
These model errors include imperfect atomic line data, poor treatment of granulation,
neglect of NLTE effects, the exact shape of the instrumental profile, etc.

From version 433 onwards, SME provides an alternative parameter uncertainty estimate
that assumes model errors dominate, rather than measurement error.
There are many ways to assess model error, none perfect.
We begin by identifying pixels that are sensitive to small changes in a free parameter.
In practice, we select all unmasked pixels (with index $i$)
where the partial derivative of the synthetic spectrum
with respect to the free parameter is nonzero (${\partial F_i}/{\partial p}\ne0$)
and the observed minus model residual ($R_i$)
is less than 5 times the measurement uncertainty for each observed spectrum pixel.
For each sensitive pixel, we estimate the parameter change ($\Delta p_i$) needed
to make $R_i$ equal to zero, assuming that $R_i$ is a linear function of $p$.
Mathematically,
\begin{equation}
\Delta p_i = R_i / \frac{\partial F_i}{\partial p}.
\end{equation}
If model errors dominate, then the width of the distribution
of $\Delta p_i$ values is a measure of parameter uncertainty.
For normally distributed errors, the distribution is Gaussian,
but model errors need not have a Gaussian distribution.
Instead we construct $C(\Delta p$),
the normalized cumulative distribution function of $\Delta p_i$ values.
Then $\sigma_\mathrm{lo}=\Delta p$ where $C(\Delta p)=0.16$,
$\sigma_\mathrm{hi}=\Delta p$ where $C(\Delta p)=0.84$,
and the mean is $\sigma=(\sigma_\mathrm{lo}+\sigma_\mathrm{hi})/2$.
This cumulative distribution function approach is robust against large residuals,
as long as they comprise fewer than 15\% of the points at either extreme.
The extent of the central part of the cumulative distribution function
provides a measure of model uncertainties.
In many cases the actual parameter uncertainty is approximately bounded
by the formal uncertainty (from the covariance matrix) 
and the model uncertainty.

Figure~\ref{parameter uncertainties} shows examples of normalized cumulative distributions.
Note that the estimated median values do not exactly match the SME result.
The reason is that the error estimate procedure
presented here is not taking into account the cross-talk between parameters. This is well
illustrated in Figure~\ref{chisq surface} showing the slice of the $\chi^2$ space along the
 \teff-\logg\ plane. Double-line contours show the uncertainties of the numerical procedure
 (small circle in the center) and the new uncertainty estimates for the effective temperature
 and surface gravity that attempt to account for the deficiencies of the model. The thick lines
 show the constant value contours of the $\chi^2$ surface. While the new estimate is
 very close (in size) to the true 1 $\sigma$ contour it does not account for
interdependence of the two parameters, as demonstrated by the difference in orientation.
When using SME to derive stellar parameters, it is imperative to understand
the underlying physics and provide balanced and (as much as possible)
orthogonal constraints for the free parameters.

\section{Miscellaneous and coming features}\label{miscellaneous features}

One of ``undocumented'' capabilities of SME is the direct calling of the library functions. Such calls
could be useful for getting more information about the final model that is not saved in the output
structure (e.g., partial number densities of various species, continuous and line opacities, etc.).
Alternatively, direct calls can be used for other purposes, such as computing monochromatic
optical depth. The SME documentation explains now how to make direct library calls and
the distribution includes examples of such calls from IDL for solving the equation of state and
continuous opacity at specified wavelengths.

Another useful tool included in the latest distribution simplifies porting a mask from
one SME structure to another. Such need often comes up when analysing a group of similar
targets using the same spectral intervals. Both structures should include observations and radial
velocities. The tool called \verb+port_mask+ takes care of resampling and interpolating the mask.
It can work on structures loaded in IDL session or stored in files.

We have successfully tested a new radiative transfer solver, that combines the current attenuation
operator algorithm with Gauss-Seidel iterations to account for scattering. Proper inclusion of
scattering is critically important for low-absorption environments such as the atmospheres of 
metal-poor giants. At moment of writing, the new solver is still in the testing phase, but we
expect it to be included in SME in the near future. 

\section{Concluding remarks}

The current version of the SME incorporates all the capabilities needed to determine
stellar parameters and chemical composition for a broad range of stars, including cool
dwarfs and giants. The package includes an equation of state that solves for chemical
equilibrium, a faster and more accurate radiative transfer algorithm, new grids of model
atmospheres, and grids of pre-computed departure coefficients for key elements.
While useful, these tools do not replace the insight of a trained researcher.
The SME user must still obtain and reduce the observations, select spectral regions,
adjust the mask, collect atomic and molecular data, and interpret the final results.

\begin{acknowledgements}
The authors are very grateful to Ulrike Heiter for writing and maintaining SME documentation,
to Thomas Nordlander for contributing NLTE tools and testing the model atmosphere grid interpolation,
and to Paul Barklem and Remo Collet for contributing the partition functions for atoms, molecules
and negative ions as part of our collaboration on the EOS package. We would also like to thank
all SME users for continuous and useful feedback on the package.
\end{acknowledgements}

%
%

\end{document}